\documentclass{article}

\usepackage{spconf}
\usepackage{amsfonts,amssymb,amsmath,amsthm,amstext}
\usepackage{graphicx,url}
\graphicspath{{./Figures/}}

\usepackage{pdfsync}
\usepackage{epstopdf}

\usepackage{mathtools}
\usepackage{subfigure}
\usepackage{float}

\usepackage[sort&compress, numbers]{natbib}

\usepackage[footnotesize]{caption}

\usepackage[textwidth=1.3cm,textsize=tiny]{todonotes}
\newcommand{\Rbb}{\mathbb{R}}

\newcommand{\scp}[2]{\langle #1, #2 \rangle}

\newcommand{\tinv}[1]{{\textstyle\frac{1}{#1}}}
\newcommand{\half}{\tinv{2}}
 
\renewcommand{\leq}{\leqslant}
\renewcommand{\geq}{\geqslant}
\renewcommand{\tilde}{\widetilde}

\DeclareMathOperator{\st}{\ {\rm s.\!t.}\ }

\DeclareMathOperator{\sign}{sign}
\DeclareMathOperator{\supp}{supp}

\DeclareMathOperator{\iSNR}{iSNR}
\DeclareMathOperator{\oSNR}{oSNR}
\DeclareMathOperator{\dB}{dB}

\DeclareMathOperator{\Id}{\mathbb{I}\hspace{-1.24mm}\mathbb{I}}
\DeclareMathOperator*{\argmin}{\arg\min}
\newcommand{\bs}{\boldsymbol}
\newcommand{\cl}{\mathcal}
\newcommand{\ie}{\emph{i.e.}, }
\newcommand{\eg}{\emph{e.g.}, }

\newcommand{\sq}{\vspace{-2mm}}

\newcommand{\C}[1]{}

\title{\vspace{-12mm} Consistent Iterative Hard Thresholding for Signal DeClipping\sq}

\name{S. Kitic$^1$, L. Jacques$^1$, N. Madhu$^2$, M. P. Hopwood$^2$, A. Spriet$^2$ and C. De Vleeschouwer$^1$.\sq
\thanks{LJ and CDV are supported by the Belgian FRS-FNRS fund. NM, MPH
  and AS are supported by NXP software, Leuven. Part of this work has
  been funded by the SPORTIC project (WIST3), Walloon Region, Belgium.}}
\address{\ninept $^1$ELEN Departement, ICTEAM, Universit\'{e} catholique de Louvain, Belgium.\\
\ninept $^2$NXP Software, Leuven, Belgium.\sq
}

\begin{document}
\maketitle

\begin{abstract}
Clipping or saturation in audio signals is a very common problem in signal processing, for which, in the severe case, there is still no satisfactory solution. In such case, there is a tremendous loss of information, and traditional methods fail to appropriately recover the signal. We propose a novel approach for this signal restoration problem based on the framework of Iterative Hard Thresholding. This approach, which enforces the consistency of the reconstructed signal with the clipped observations, shows superior performance in comparison to the state-of-the-art declipping algorithms. This is confirmed on synthetic and on actual high-dimensional audio data processing, both on SNR and on subjective user listening evaluations.
\end{abstract}
\begin{keywords}
Signal Clipping, Sparse Recovery, Inverse Problems, Greedy Methods, Audio Processing.
\end{keywords}

\sq
\section{Introduction}
\label{sec:intro}
\sq
Signal \emph{clipping} is the corruption of the dynamic range of a signal, manifested as a corruption of the signal magnitude at some boundary level. This phenomenon usually occurs during the very first stages of signal recording, typically when the input range of a device is not sufficiently large as in A/D converters or when the response of a system is not linear beyond a certain level. Quite naturally, the task of ``declipping'' such a corrupted signal has therefore attracted a lot of research recently in the signal processing community, and in particular, in the audio restoration field -- bearing in mind the sensitivity of the human ear to unnatural sound artifacts.  

\begin{figure}
  \centering
  \includegraphics[width=0.75\columnwidth]{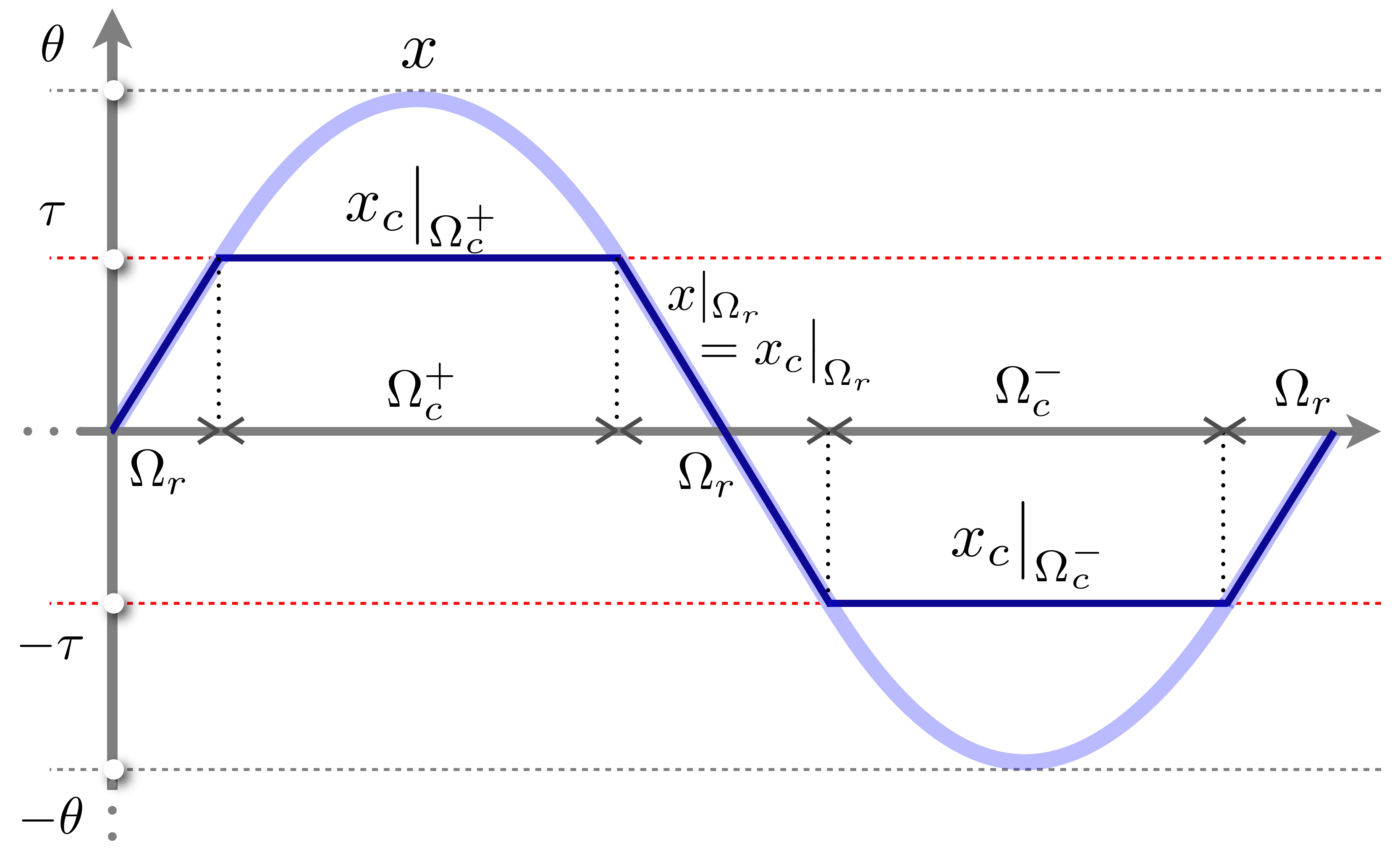}
  \caption{Hard-clipping example: $x=x(t)$ is the original signal
    (light curve) and $x_c(t)$ is its clipped version
    (dark curve). The notations are explained in~Sec.~\ref{sec:clipping-model}.\sq \sq}
  \label{Clipping}  
\end{figure}

There exist actually two main kinds of clipping models: hard and soft clipping. The first, as illustrated in Fig.~\ref{Clipping}, simply replaces the saturated signal amplitudes by some constant saturation level, while the second, which is not treated in this paper, corresponds to a reduction of the amplitude gain beyond this level.
Once a clipped audio signal is replayed, humans perceive it as an unnatural and unpleasantly distorted sound. In music, for instance, all notes of a clipped sound seem loud, because both soft and loud notes are clipped to the same level, which reduces the auditive contrast.

This paper presents a novel iterative method for signal
\emph{declipping}. As explained in Sec.~\ref{sec:sparse}, this is based on the premise that the underlying signal has a \textit{sparse} structure in a convenient representation. After introducing a model for the clipping scenario in Sec.~\ref{sec:clipping-model}, Sec.~\ref{sec:decl-inverse-probl} defines the declipping operation as an inverse problem regularized by the sparsity assumption and which stays consistent with the whole clipping process. The main interest of this formulation is to provide guidelines for developing a \emph{consistent} variant of the Iterative Hard Thresholding \cite{IHT2004} adapted to the non-linear clipping alteration. After a brief review of the literature in the field (Sec.~\ref{sec:StateOfTheArt}), the efficiency of this approach is finally demonstrated on synthetic and actual audio signal restoration in comparison with state-of-the-art methods (Sec.~\ref{sec:Benchmarks}).  

\noindent\emph{Conventions:} Most of domain dimensions (\eg $M$, $N$) are denoted by capital roman letters. Vectors and matrices are associated to bold symbols while lowercase light letters are associated to scalar values.  The $i^{\rm th}$ component of a vector $\bs u$ is $u_i$ or $(\bs u)_i$. The identity matrix is $\bs\Id$. Vectors of zeros and ones are denoted by $\bs 0$ and $\bs 1$ respectively.  The set of indices in $\Rbb^D$ is $[D]=\{1,\,\cdots,D\}$. Scalar product between two vectors $\bs u,\bs v \in \Rbb^{D}$ reads $\bs u^T \bs v = \scp{\bs u}{\bs v}$ (using the transposition $(\cdot)^T$). For any $p\geq 1$, $\|\cdot\|_p$ represents the $\ell_p$-norm such that $\|\bs u\|_p^p = \sum_i |u_i|^p$ with $\|\bs u\|=\|\bs u\|_2$ and $\|\bs u\|_\infty = \max_i |u_i|$. The $\ell_0$ ``norm'' is $\|\bs u\|_0 = \# \supp \bs u$, where $\#$ is the cardinality operator and $\supp \bs u = \{i: u_i \neq 0\} \subset [D]$.  For $\cl S \subset [D]$, $\bs u|_{\cl S}\in \Rbb^{\#\cl S}$ (or $\bs \Phi|_{\cl S}$) denotes the vector (resp. the matrix) obtained by retaining the components (resp. columns) of $\bs u\in\Rbb^D$ (resp. $\bs \Phi\in\Rbb^{D'\times D}$) belonging to $\cl S\subset [D]$. Alternatively, $\bs u|_{\cl S} = \bs R_{\cl S} \bs u$ or $\bs \Phi|_{\cl S} = \bs \Phi \bs R^T_{\cl S} $ where $\bs R_{\cl S} := (\Id|_{\cl S})^T \in \{0,1\}^{\#\cl S \times D}$ is the \emph{restriction} operator. We denote by $(\bs x)_+$ the positive thresholding $(x_i)_+ = (x_i + |x_i|)/2$, while the negative counterpart reads $(\bs x)_- = -(-\bs x)_+$. 

\sq
\section{Sparse Signal Representation}
\label{sec:sparse}
\sq 
The vast majority of real-life audio signals have compressible structures, meaning that these signals may be represented or approximated as the linear combination of few elements taken in a set of elementary wave forms (\eg DCT, Wavelets \cite{plumbley2010sparse,mallat09wavelet}). 

Mathematically, this \emph{sparsity} concept is applied to 1-D temporal signals (\eg audio) as follows. We assume that, within a certain time window $T\subset \Rbb$, a continuous signal $x(t)$ has been sampled with $N$ regular samples gathered in a column vector $\bs x\in\Rbb^N$. We consider then that for an appropriate sparsity basis $\bs\Psi \in \Rbb^{N\times D}$ with $N \leq D$, $\bs x$ can be described as
\sq  
\begin{equation}\label{SparseFormula}
  \bs{x} \approx\bs\Psi \bs \alpha, \text{with }\|\bs \alpha\|_0\ll N\text{ and }\|\bs x - \bs\Psi\bs \alpha\| \ll \|\bs x\|. \sq
\end{equation}
When $\bs \Psi$ is an orthonormal basis with $D=N$, there exists only one vector $\bs \alpha^*$ satisfying $\bs x=\bs \Psi\bs \alpha^*$. A sparse coefficient vector answering the problem \eqref{SparseFormula} is found by taking the best $K$-term approximation of $\bs \alpha^*$ given a fixed sparsity level $K\ll N$. In other words, $\bs \alpha=\bs \alpha^*_K=\cl H_K(\bs \alpha^*)$ where $\cl H_K$ is the $K$-term thresholding operator setting to zero all but the $K$ highest-magnitude coefficients of $\bs\alpha^*$. If $D > N$, there are many coefficient vectors whose re-synthesis with $\bs\Psi$ approximates $\bs x$. This redundancy is often useful to select sparser coefficient vector. Despite the NP-hardness of \eqref{SparseFormula} \cite{SparseNP}, ``relaxed'' optimization methods and greedy algorithms exist in order to find such sparse vectors under additional requirement on $\bs\Psi$ \cite{BP,GreedIsGood}.

Amongst them, the Iterative Hard Thresholding (IHT) offers interesting advantages like fast convergence and provable sparse decomposition guarantees \cite{IHT2004}. If a signal $\bs x$ is assumed $K$-sparse in $\bs \Psi\in\Rbb^{N\times D}$ (with $K\ll N$), this algorithm is designed to find one minimizer of a Lasso-type \cite{tibshirani1996regression} restatement of~(\ref{SparseFormula}):\sq
\begin{equation}
\label{Lasso_IHT}
\hat{\bs\alpha} = \argmin_{\tilde{\bs\alpha}\in\Rbb^D} \half \|\bs x - \bs\Psi\tilde{\bs\alpha}\|^2 \st \|\tilde{\bs \alpha}\|_0 \leq K. \sq
\end{equation}
IHT approximates the solution of \eqref{Lasso_IHT} by performing the following iterative evaluation:\sq
\begin{equation}
\label{IHT_K} 
\bs\alpha^{(n+1)}=\cl H_K[\bs\alpha^{(n)}+\bs\Psi^T(\bs x - \bs\Psi\bs\alpha^{(n)})],\ \bs\alpha^{(0)}=\bs 0.
\end{equation}
In words, at each iteration, this algorithm hard-thresholds the previous solution updated by a gradient descent on the fidelity cost of \eqref{Lasso_IHT}. 

The IHT procedure is very general and is also applicable to the
recovery of signals indirectly observed by a sensing matrix
$\bs\Phi\in\Rbb^{M\times N}$, as in the Compressed Sensing (CS) framework \cite{CSintro,IHTforCS}. In such case, the algorithm above simply undergoes the replacement $\bs\Psi \to \bs\Phi\bs\Psi\in\Rbb^{M\times D}$ for integrating this sensing.
 
\sq
\section{Clipping Model}
\label{sec:clipping-model}
\sq
Assuming a symmetric clipping (as in Fig.~\ref{Clipping}) associated to a clipping threshold $\tau>0$, the (hard) clipping operation $C_\tau$ is mathematically defined as:\sq
\begin{equation}
\label{eq:clipping_operator}
\bs x_c = C_\tau(\bs x) := \min(|\bs x|, \tau) \sign(\bs x),\sq
\end{equation}
where all the operations are applied component wise on $\bs x$.

From this clipping operation, we can actually define different sets of samples in $\bs x$. The set of reliable data, those which are not subject to clipping, is $\Omega_r=\{i\in[N]: |x_i|< \tau\}$, while the clipped index set $\Omega_c=\{i\in [N]: |x_i|\geq \tau\}$ can be split into two disjoint subsets $\Omega^\pm_c=\{i\in [N]: \pm\,x_i \geq \tau\}$, with $\Omega_c=\Omega^+_c\cup\Omega^-_c$ and $\Omega_r\cup\Omega_c=[N]$. 

In this work, we assume that these sets are known. This happens for instance if the clipping process is hard and not corrupted by a strong noise, in which case all the previous sets can be deduced from the observation of $\bs x_c$. 

The knowledge of the sets $\Omega_r$ and $\Omega_c^\pm$ simplifies the corresponding forward model (\ref{eq:clipping_operator}), \ie \sq
\begin{equation}
  \label{eq:linearized_clipping}
  \bs x_c = \bs M_{\Omega_r} \bs x + \tau \bs M_{\Omega^+_c} \bs 1 - \tau \bs M_{\Omega^-_c}^T \bs 1,\sq
\end{equation}
with $\bs M_{\cl S} = \bs R_{\cl S}^T \bs R_{\cl S} \in \Rbb^{N\times N}$ is a diagonal masking matrix, \ie $(\bs M_{\cl S} \bs u)_i=u_i$ if $i\in\cl S\subset [N]$ and 0 otherwise. 

\sq
\section{DeClipping Inverse Problem}
\label{sec:decl-inverse-probl}
\sq

A naive approach to the declipping problem is to use directly the IHT algorithm \eqref{IHT_K} on the partial observation of the unclipped signal samples, \ie considering $\bs{x}_c|_{\Omega_r} = \bs R_{\Omega_r} \bs x$ as the partial observations of $\bs x$ realized by the sensing matrix $\bs \Phi = \bs R_{\Omega_r} \bs\Psi$. As explained in \cite{cOMP,SparseLand}, the downfall of this method is that it does not take into account the information contained in the clipped samples, namely the clipping threshold ($\tau$) and (possibly) the uppermost absolute magnitude ($\theta$).

Therefore, in this work, inspired by the ideal objective \eqref{Lasso_IHT} targeted by IHT, we address the following inverse problem \sq 
\begin{equation}
  \label{IdealDeClippingProblem}
  \hat{\bs\alpha} = \argmin_{\tilde{\bs\alpha}\in\Rbb^D} \half\| \cl B(\bs x_c - \bs \Psi\tilde{\bs \alpha}) \|^2 \st \|\tilde{\bs \alpha}\|_0 \leq K, \sq
\end{equation}
where the corresponding reconstructed signal is $\hat{\bs{x}}=\bs \Psi\hat{\bs \alpha}$.

The key function $\cl B:\Rbb^N\to\Rbb^N$ involved in \eqref{IdealDeClippingProblem} reads \sq
\begin{equation}
  \label{EqPenal}
  \cl B(\bs u) = \bs M_{\Omega_r} \bs u + (\bs M_{\Omega_c^+} \bs u)_+ + (\bs M_{\Omega_c^-} \bs u)_-. \sq
\end{equation}
Minimizing $\cl E(\tilde{\bs x}) := \half \|\cl B(\bs x_c - \tilde{\bs x})\|^2$ forces a signal candidate $\tilde{\bs x} = \bs\Psi \tilde{\bs\alpha}$ to be \emph{consistent} with the observed clipped signal $\bs x_c$.  Indeed, from \eqref{EqPenal}, this cost can be split in three parts, \ie \sq
\begin{multline}
\label{eq:cost-declip}
\cl E(\tilde{\bs x}) := \half \| \bs M_{\Omega_r} (\bs x_c - \tilde{\bs x})\|^2\\ + \half \| \bs M_{\Omega_c^+} (\tau\bs 1 - \tilde{\bs x})_+\|^2 + \half \| \bs M_{\Omega_c^-} (-\tau\bs 1 - \tilde{\bs x})_-\|^2. \sq
\end{multline}
A small first term promotes the candidate signal to match the observed clipped signal in the unclipped domain, while, thanks to the $+$ or $-$ thresholding functions, having a minimal second (or third) term enforces $\tilde{\bs x}$ to be bigger (resp. smaller) than $\tau$ (resp. $-\tau$) on the set $\Omega_c^+$ (resp. $\Omega_c^-$).

A few remarks can be made on \eqref{IdealDeClippingProblem}. First,
this declipping program corresponds intuitively to picking from the set of all signals consistent with the clipped observation $\bs x_c$, one which has a sparsity smaller than $K$. Provided the original signal $\bs x$ (approximately) respects this sparsity requirement, the declipping program will have a solution. 

Second, we impose actually a strict sparsity model on $\tilde{\bs \alpha}$ parametrized by a sparsity order $K\ll N$ assumed optimal. We will see later how we can estimate this value. Notice also that additional constraints can be added, such as the knowledge that the original signal has a bound amplitude (see Fig.~\ref{Clipping}), \ie we can additionally impose $\|\bs\Psi \tilde{\bs \alpha}\|_\infty \leq \theta$ (\eg as done in \cite{cOMP,SparseLand}).   
\medskip 

Of course, directly solving \eqref{IdealDeClippingProblem} is as hard as recovering a sparse signal in \eqref{Lasso_IHT}. However, a novel iterative hard thresholding adjusted to \eqref{IdealDeClippingProblem} can be obtained by following the same guidelines than those used to derive IHT from \eqref{Lasso_IHT} \cite{IHT2004}. Since $(\cdot)_+^2$ is differentiable, the cost $\cl E(\bs\Psi\bs\beta)$ is actually a smooth convex function of $\bs\beta\in\Rbb^D$ whose gradient reads:
$$
\bs\nabla_{\bs\beta}\,\cl E(\bs \Psi \bs\beta) = - \bs \Psi^T \cl B(\bs x_c - \bs \Psi\bs\beta). \sq
$$
 
Consequently, remembering that the internal part of the thresholding in \eqref{IHT_K} is a gradient descent, we propose to solve a version of Iterative Hard Thresholding adjusted to DeClipping that we call IHT-DC:
\begin{equation}
  \bs{\alpha}^{(n+1)}=\cl H_K[ \bs{\alpha}^{(n)} + \mu^{(n)}\,\bs{\Psi}^T \cl B(\bs{x}_c-\bs{\Psi\alpha}^{(n)})], 
\end{equation}
where $\bs\alpha^{(0)}=\bs 0$. The value $\mu^{(n)}$ is simply selected by a fast 1-D convex minimization (\eg using golden section \cite{GoldenSection}) of\sq 
$$
g(\mu) = \cl E\big(\bs{\alpha}^{(n)} + \mu\, \bs{\Psi}^T \cl B(\bs{x}_c-\bs{\Psi\alpha}^{(n)})\big).\sq\sq\sq
$$
\sq
\section{Prior works in the field}
\label{sec:StateOfTheArt}
\sq
Different strategies have been developed so far in the literature to address the declipping problem. One of the oldest is the Autoregressive (AR) method \cite{Janssen}. AR assumes that the underlying signal can be modeled as an autoregressive process \cite{DSP}, and based on that premise it estimates the AR coefficients and interpolates the missing samples. AR relies thus on one specific generative signal model which is well adapted to speech signals only. It suffers a lack of flexibility for modeling and restoring other kind of signals (as music); an issue solved by sparsity-driven methods. 

A more recent approach is the Constrained Orthogonal Matching Pursuit (cOMP) audio inpainting algorithm \cite{cOMP}. This one is fundamentally based on Orthogonal Matching Pursuit \cite{OMP} and constrained optimization. In the first stage, cOMP discards non-reliable samples from the data and attempts to detect the optimal basis vectors using only reliable samples. In the second stage, clipping constraints are imposed to the chosen basis using some external optimization toolbox. Despite reported improvements relatively to the AR declipping results, one drawback of this algorithm is that the first stage does not take into account the information stored in the clipped samples. This may yield to an incorrect sparse support estimation impacting the whole method. A second drawback is that the number of iterations is directly related to the length of the estimated sparsity support, due to the fact that OMP gradually increases the support length at each iteration.

\ \\[-8mm]
\indent The work in \cite{SparseLand} solves a variant of \eqref{IdealDeClippingProblem} where the sparsity inducing $\ell_0$ ``norm'' is minimized and the clipping consistency is a constraint. This technique uses a reweighted $\ell_1$ minimization for approaching $\ell_0$ \cite{rwl1} and it addresses the shortfall of the cOMP. Unfortunately, we were unable to extensively run the corresponding toolbox in our high dimensional audio setting (see Sec.~\ref{sec:Benchmarks}). There exist also some commercial declipping softwares like the Adobe\textsuperscript{\textregistered} Audition DeClipper\textsuperscript{\texttrademark}. However, their black box nature make difficult any theoretical comparison with other methods.

Finally, but not directly related, clipping is associated to saturation in compressive data quantization \cite{laska2011democracy} or to time-frequency signal corruption \cite{Smaragdis}, while in the extreme case where $\tau\to 0^+$, the cost $\cl E$ in~\eqref{eq:cost-declip} is reminiscent of the energy implicitly minimized in the Binary Iterative Hard Thresholding (BIHT) in the context of 1-bit CS~\cite{jacques2011robust}.\sq 

\sq
\section{Experiments}
\label{sec:Benchmarks}
\sq\sq
\begin{figure}[h]
	\centering
        \includegraphics[width=0.75\columnwidth]{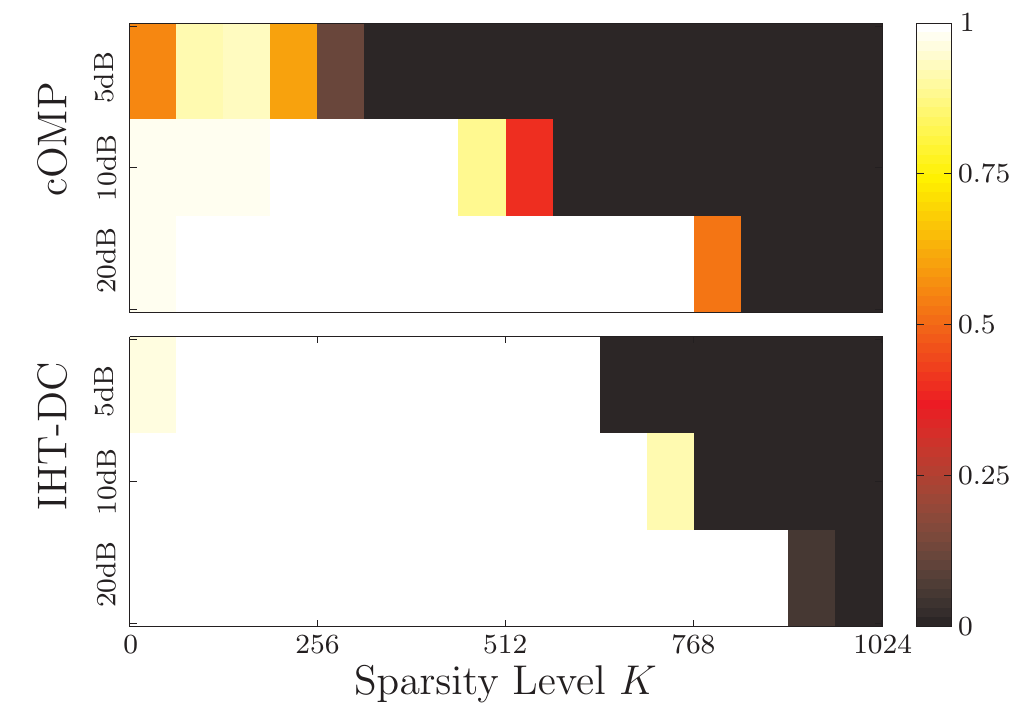}\sq
\caption{Probability of success for accurate recovery under severe ($5\dB$), moderate ($10\dB$) and mild ($20\dB$) clipping. \sq\sq}
        \label{Synthetic}
\end{figure}

Two sets of the benchmark tests have been conducted, one with synthetic signals and the other with actual audio (music) data. For both data types, signals are degraded by hard-clipping \eqref{eq:clipping_operator}. The performance of the algorithms are measured in dB through input and output (restored signal) SNRs. These metrics are defined as $\iSNR = 10 \log_{10}(\| \bs x\|^2/\|\bs x-\bs x_c\|^2)$ and $\oSNR=10 \log_{10}(\|\bs x\|^2/\| \bs x -\hat{\bs x}\|^2)$ respectively, where $\bs x$, $\bs x_c$ and $\hat{\bs x}$ are the original, the clipped and the declipped signals, respectively.

\vspace{1mm}
\noindent\textbf{Benchmark on synthetic data:} This type of benchmark is suitable for sparsity-based declipping algorithms such as cOMP and IHT-DC.  Synthetic signals have been generated in $\Rbb^{N=1024}$ as random $K$-sparse signals in a DCT basis $\bs \Psi\in\Rbb^{N\times N}$ for various values of $1\leq K \leq N$. These sparse signal were defined by first selecting uniformly at random a $K$-length support $T$ in $[D]$ and by drawing identically and independently the dictionary coefficients of this support according to a normal distribution, the coefficients outside of $T$ being set to 0.

For both reconstruction methods, the ``success'' of a declipped reconstruction has been arbitrarily defined as $\oSNR>80\dB$. For different values of $K$ and for different clipping scenarios, we estimated the probability of the accurate recovery by the frequency of success over 100 trials. The results for 3 levels of clipping are presented in Fig.~\ref{Synthetic}. These are defined as ``mild'' ($\iSNR\!=\!20\dB$), ``moderate'' ($\iSNR\!=\!10\dB$) and ``severe'' ($\iSNR\!=\!5\dB$). It appears that the proposed method outperforms cOMP in all cases. The advantage of using IHT-DC is especially pronounced for the severe and moderate clipping, while it becomes smaller for the mild clipping level. This is in accordance with the fact that, at this level, more reliable observations are available and cOMP manages to detect the correct basis vectors more often. Interestingly, the global appearance of Fig.~\ref{Synthetic} is similar to the transition recovery diagram existing in sparse signal recovery in Compressed Sensing \cite{PhaseTransitions}, the $y$-axis in this figure measuring implicitly the amount of reliable observations.

About the processing time, our programming of IHT-DC in Matlab$^\copyright$ is fast when the sparsity level is smaller than the success/failure transition point reported in Fig.~\ref{Synthetic}. For instance, under moderate clipping and for $K\leq 512$ (the transition occurring in $K\in[576,768]$), the recovery is perfect and IHT-DC stops after less than 100 iterations, \ie less than 2s on a 2.93GHz laptop. 

\vspace{1mm}
\noindent\textbf{Benchmark on audio data:} In this benchmark, we test
the efficiency of our approach on two actual samples of music track,
\ie``You Are My Kind'' by \textit{Seal \& Santana} (15s) and ``Sultans
of Swing'' by \textit{Dire Straits} (10s), each one having a very
different speech and tonal content. Each track is sampled at 16kHz
(32bits per sample), and the declipping operation was realized by
individual processing of 75\%-overlapping temporal slices of length
$N=1024$, before to re-synthesize the full length music signal with a
symmetric sinusoidal window \cite{cOMP}. 

Since the audio signals are not truly sparse, IHT-DC cannot be readily applied to restore them in a clipping scenario. In order to avoid us to arbitrarily set one sparsity level $K$, \eg according to the class of audio signals considered, we prefer to make the IHT-DC adaptive by enforcing it to ``learn'' an optimal sparsity level $K$. 
Selecting $\bs\Psi$ as a two-times redundant DCT dictionary for efficient audio signal representations \cite{plumbley2010sparse}, this adaptivity is achieved by gradually increasing the sparsity requirement by~1 at each iteration, starting from a low value, until the residual between the clipped observations and the reclipped reconstructed signal has a sufficiently small energy. 

The \emph{SNR gain} of this adaptive IHT-DC is then measured by the difference between oSNR and iSNR; obviously, it should be as high as possible. The values in Table~\ref{NumEvalMusic}(top) show the SNR gains for different clipping levels illustrating severe ($5\dB$), moderate ($10\dB$) and mild ($15\dB$) signal corruption, in comparison with other methods. Once again, IHT-DC outperforms all others in this benchmark. Furthermore, only Adobe software (apart from the proposed algorithm) is somewhat successful in improving the clipped signal, whereas other methods usually degrade signals even more. Overall, IHT-DC was about 5 times faster than cOMP, with ratio between processing time and track duration close to 60 and 240 for moderate and hard clipping, respectively. 

\begin{table}
  \newcommand{\ph}{\phantom{-}}
  \centering
  \ninept
  \begin{tabular}{|@{\ }r@{\ }|@{\,}c@{\,}|@{\,}c@{\,}|@{\,}c@{\,}|@{\,}c@{\,}|@{\,}c@{\,}|@{\,}c@{\,}|}
    \hline
    SNR gain&\multicolumn{3}{c@{\,}|@{\,}}{Music 1}&\multicolumn{3}{@{}c@{}|}{Music 2}\\
    \hline    \scriptsize$\iSNR=$&\!\scriptsize$5\dB$\!&\!\scriptsize$10\dB$\!&\!\scriptsize$15\dB$\!&\!\scriptsize$5\dB$\!&\!\scriptsize$10\dB$\!&\!\scriptsize$15\dB$\!\\[.2mm]
   \hline
   &&&&&&\\[-3mm]
    Adobe DC&$\ph 2.9$&$\ph 4.6$&$\ph 5.1$&$\ph 0.5$&$\ph 1.2$&$\ph 2.0$\\
    AR&$-1.6$&$-2.2$&$\ph 1.6$&$-1.2$&$-0.5$&$-0.5$\\
    IHT-DC&$\bf\ph 4.4$&$\bf\ph 6.5$&$\bf\ph 7.2$&$\bf \ph 5.4$&$\bf \ph 5.9$&$\bf \ph 5.4$\\
    cOMP&$-2.4$&$-4.5$&$-2.3$&$-2.6$&$-1.7$&$-1.2$\\
    \hline
    \multicolumn{1}{c}{\ }\\[-3.5mm]
    \hline
    Evaluation&\!\scriptsize$5\dB$\!&\!\scriptsize$10\dB$\!&\!\scriptsize$15\dB$\!&\!\scriptsize$5\dB$\!&\!\scriptsize$10\dB$\!&\!\scriptsize$15\dB$\!\\[.2mm]
   \hline
   &&&&&&\\[-3mm]
  Adobe DC&$1.5$&$3.0$&$4.0$&$2.0$&$3.0$&$4.0$\\
  AR&$1.0$&$3.0$&$4.0$&$1.0$&$3.0$&$4.0$\\
  Clipped&$1.0$&$2.0$&$4.0$&$1.5$&$2.0$&$4.0$\\
  IHT-DC&$\bf 3.0$&$\bf 4.0$&$4.0$&$\bf 3.0$&$\bf 4.0$&$4.0$\\
  cOMP&$1.0$&$2.0$&$4.0$&$1.0$&$3.0$&$4.0$\\
  \hline
\end{tabular}\sq
  \caption{(top) The SNR gain (in dB) \emph{vs} clipping for the two audio tracks. (bottom) Listening evaluation (16 pers.) for the two music samples. Evaluation score are between 1 ``very poor'' and 5 ``excellent quality''. \vspace{-7mm}}
  \label{NumEvalMusic}
\end{table}

In addition to the previous SNR gain comparisons, we also wish to
depict how end-users perceive audio data processed by the different
algorithms. Therefore, we have designed a randomized subjective test
as follows. We have considered again the 6 clipped audio contents
generated by applying our 3 clipping scenarios to the 2 music
tracks. Then, 5 resulting versions for each clipped track have been
collected: the reconstructions obtained for each of the 4 declipping
algorithms, plus the raw clipped version. A blind test has then been
run independently with 16 subjects. The 6 clipped tracks have been
considered in a random order (to prevent the ``training effect''). For
each clipped track, the 5 versions were evaluated by the listeners in
a random order (preserving objectivity). The subjects were asked to score the result between 1 (very poor) and 5 (excellent).

The results are presented at the Table~\ref{NumEvalMusic}(bottom). For the two music signals the proposed algorithm clearly outperforms the competitors. Again, it seems that only Adobe DeClipper (Adobe DC) provides some improvement in perceived quality, while cOMP and AR methods are in some cases graded lower than actual clipped signal. In case of a mild clipping, users were in most cases unable to hear the difference between audio clips.\sq\sq

\section{Conclusion}
\label{sec:conclusion}
\sq
A new iterative method for declipping signals has been presented. This extends the IHT algorithm by integrating a clipping consistency during the iterations. Experimental results on synthetic and actual audio data have demonstrated the efficiency of this approach compared to other known techniques. In the future, we plan to justify the theoretical conditions for guaranteeing the convergence of the IHT-DC. In particular, we will analyze how the ``phase'' transition diagram obtained in Fig.~\ref{Synthetic} can be predicted according to both the clipping and the signal sparsity levels. On the practical perspective, an important gain could also be obtained by exploiting different dictionaries. The redundant DCT used for our tests is a very generic dictionary that is not especially well suited for purely  speech signals, for instance. 


\end{document}